\documentstyle[epsf]{elsart}
\begin{document}

\begin{frontmatter}
\title{
  The 1 Teraflops QCDSP computer
}
\author{
  ROBERT D. MAWHINNEY\thanksref{DOE}
}
\address{
  Department of Physics,
  Columbia University,
  New York, NY  10027,  USA
}
\thanks[DOE]{This work is supported in part by the United States
Department of Energy, the RIKEN-BNL Research Center and
Brookhaven National Laboratory.  The many contributors to
the work described here are listed in the Acknowledgements section.}
\begin{abstract}
  The QCDSP computer (Quantum Chromodynamics on Digital Signal
  Processors) is an inexpensive, massively parallel computer
  intended primarily for simulations in lattice gauge theory.
  Currently, two large QCDSP machines are in full-time use:
  an 8,192 processor, 0.4 Teraflops machine at Columbia University
  and an 12,288 processor, 0.6 Teraflops machine at the RIKEN-BNL
  Research Center at Brookhaven National Laboratory.  We describe
  the design process, architecture, software and current physics
  projects of these computers.
\end{abstract}
\end{frontmatter}

\def\thepage{CU--TP--966}
\thispagestyle{myheadings}

\section{Introduction}

The search for the smallest constituents of matter has led to the
discovery of many sub-atomic particles and the development of the
standard model of particle physics.  This model is based on the
principle of ``local gauge invariance", first seen in Maxwell's theory
of electromagnetism, where it constrains the types of interactions
possible between photons and electrons.  The standard model includes
the strong, weak and electromagnetic forces, providing a description of
virtually all experimental phenomenon seen to date.  It is a theory of
generalized force-carrying particles of spin one interacting with
matter that is either fermionic (spin one-half) or bosonic (spin
zero).

The principle of local gauge invariance is an important abstract idea,
similar to the concept of evolution in biology, but as embodied in the
standard model it also leads to a quantitative theory which describes
particle interactions precisely.  Many comparisons between standard
model predictions and experiment have been made, primarily involving
the weak and electromagnetic part of the model or the strong force at
high energies.  At low energies (up to a few GeV) the strong force
(described by a part of the standard model known as quantum
chromodynamics, QCD) is not analytically tractable in any reliable
approximation, making quantitative predictions in this region solely
accessible by computational techniques.  As a simple example, most of
the proton's properties are completely determined by QCD, but cannot be
calculated from first principles.

\pagenumbering{arabic}
\addtocounter{page}{1}

The lack of precise predictions from QCD is currently a restriction on
further tests of the standard model and our ability to understand new
phenomena to be probed by experiment.  QCD is formulated in terms of
quarks (spin one-half particles) and gluons, which mediate the strong
force.  The electroweak interactions of quarks are given by the
standard model, but many of the manifestations of these interactions
are only visible in physical particles (hadrons) which are bound states
of quarks.  Precise predictions from QCD for electroweak processes
requires knowing precise information about the quark content of
hadrons.  In addition, the RHIC (Relativistic Heavy Ion Collider) at
Brookhaven National Laboratory will soon begin colliding nuclei at high
energies to probe nuclear matter at high temperatures and densities.
Once again analytic calculations here are limited, even though the
underlying physical formulation is presumed understood.

\section{Lattice Gauge Theory}

The need for accurate calculational results from QCD is vital for many
areas of research in particle physics.  QCD is also of intrinsic
interest as a theory in its own right and as a prototype for other,
more fundamental theories of nature.  For almost 20 years, QCD has been
the subject of numerical investigations.  As a calculational problem,
QCD is particularly straightforward, since only a few free parameters
(the strength of the strong force at some distance scale and masses for
the quarks) completely define the theory.  It is, however, very
computationally intensive.

When QCD is formulated on a space-time grid, it is generally referred
to as lattice QCD \cite{lgt_intro}.  Most numerical work on QCD uses the
Feynman path integral approach, where the quantum mechanical nature of
the system is exhibited by summing over all possible configurations of
quarks and gluons, weighted by the classical action for such a
configuration.  In this sum over configurations, we measure values for
various observables, which are related to the physical quantities of
interest.  To evaluate an observable $O_i$ we must calculate a
multi-dimensional integral \begin{equation}
  \langle 0_i \rangle = \frac
    { \int \prod_{n, \mu} \; dU(n, \mu) \det\{ D_{([U],m)} \}
	  \exp( - \beta S[U]/\hbar ) \; O_i } { \int \prod_{n, \mu} \;
    dU(n, \mu) \det\{ D_{([U],m)} \}
      \exp( - \beta S[U]/\hbar ) }
\end{equation}
where $n$ runs over all space-time points, $\mu = 0$ to 3 runs over the
4 directions in space-time, $U$ is an SU(3) matrix, $dU$ is the gauge
invariant Haar measure on the group SU(3), $D_{([U],m)}$ is one of the
possible lattice Dirac operators, $m$ is the quark mass, $S[U]$ is the
classical action for a gauge field $U$, $\beta$ is the inverse of the
couping constant squared and $\hbar$ is Planck's constant.

As written, the path integral over the matrices $U$ represents the
integration over all gluon degrees of freedom.  The quark integrations
have already been done and their effects are included through the
determinant factor above.  The largest part of the computational
load in lattice QCD comes from evaluating this determinant for a
fixed background gauge field $U$.

The lattice Dirac operator $D$ is a linear operator, which depends on
$U$, and has a dimensionality greater than the number of space-time
points.  (A $32^3 \times 64 $ space-time volume contains $10^7$
points.)  Discretizing the continuum Dirac operator for lattice
simulations produces a variety of different lattice operators which
should all give the same physics back in the continuum limit.  Common
lattice Dirac operators are the Wilson \cite{lgt_intro},
staggered \cite{lgt_intro}, domain wall \cite{shamir} and
overlap/Neuberger \cite{n-n} operators.  The first two are in wide use
by the community, the third will be described more later in this
article and the fourth is described in another paper in this series.

The importance sampling algorithms (hybrid molecular dynamics and
hybrid Monte Carlo) generally employed in lattice QCD do not require a
calculation of the determinant.  By writing the determinant as the
exponential of the trace of the logarithm of the matrix, only the trace
of the Dirac operator is needed.  The trace of the Dirac operator is in
turn found with a a stochastic estimator, which means solving a linear
system involving the Dirac (quark) matrix.  This matrix is large, but
sparse, only of $O(10)$ non-zero entries per row or column.  It is easy
to parallelize this linear equation problem, since the data flow is
regular and known.  On each local processor, one must be able to
efficiently multiply $3 \times 3$ complex matrices with a complex
3-vector.

Ultimately, better algorithms may be found, but the presence of
fermions has hindered progress in this area for many years.  Given the
importance of QCD calculations in relating experimental results to
standard model parameters, an obvious way to proceed is to develop
inexpensive, scalable, massively parallel computers to run lattice
QCD.  This approach has been pursued for many years by a number of
groups, including the group at Columbia.  We now turn to the most
recent machine developed primarily at Columbia, QCDSP (Quantum
Chromodynamics on Digital Signal Processors).

\section{QCDSP} 

The design of the QCDSP computer began in the spring of
1993 \cite{nhc}.  At that time, a number of dedicated, special purpose
QCD computers were in operation \cite{iwasaki} (including ACP-MAPS at
Fermilab, GF11 at IBM, APE in Italy, QCD-PAX in Tsukuba and the
Columbia 256 node machine).  Sustained speeds of $\sim 5$ Gigaflops
were being achieved, but most of the simulations were studying quenched
QCD, i.e.\ QCD with the determinant factor in the path integral set to
1.  This is an uncontrolled approximation to the full theory, which
would require computers on the Teraflops scale to remove.

Costs for a general purpose commercial teraflops scale machine at this
time were estimated at around \$100 million US dollars, for delivery in
two years.  A joint commercial/academic effort in the US estimated a
few tens of millions of US dollars for a commercial computer slightly
customized to run QCD at a Teraflops scale \cite{rdm_lat94}.  This
project never materialized, with one cause being the large cost.  The
very successful CP-PACS project in Tsukuba, Japan followed a
commercial/academic path, culminating in the 600 Gflop Hitachi computer
and is described elsewhere in this volume.

The major design goal of QCDSP was maximum sustained performance for
QCD per unit cost in a machine which could scale to a peak performance
of at least a Teraflops.  Initial estimates, detailed below, gave a
price for parts of 3 million US dollars.  Low costs required inexpensive
processors and the Teraflops goal required the machine to scale to very
large numbers of processors (10,000 or more).  Another important part
of cost effectiveness is the ratio of money spent on processors, memory
and communications hardware.  Since the communications patterns for the
currently preferred QCD algorithms are very regular and dominantly
involve transfer of data between nearest neighbor space-time points, a
machine with a grid based communications network works very well for
QCD.

A grid based network is quite straightforward to design and inexpensive
to build.  Since no routing information is needed for nearest neighbor
transfers (the hardware directly connects nearest neighbors) the
network has very little startup latency.  This is important if most
data transfers (as is the case for QCD on this type of machine) involve
sending many small amounts of data.  The nearest neighbor grid
architecture of course allows general communications to be done by
hopping data between processors, which requires routing information and
decreases network bandwidth.

A four dimensional grid-based communication network was chosen for
QCDSP.  This is natural, since our problem is one in four-dimensional
space-time (although the domain wall fermions described below are
naturally thought of in five dimensions).  This makes the mapping of
the problem to the machine particularly simple; each processor is
responsible for the data storage of variables for a particular
space-time volume.  The communication between processors is then
dominantly nearest neighbor (except for global sums which are
described in more detail below).  One can always run a four dimensional
problem on a lower dimensional regular grid (still using the natural
mapping) by making some dimensions local to the processor.

Another advantage of the four-dimensional nearest neighbor
communications network is that no single dimension need be large for a
machine with a very large number of processors.  For example, 10,000
processors are contained in a four-dimensional hypercubic lattice with
10 processors in each dimension.  Since the natural mapping described
above implies that the size of the physics problem in each dimension is
greater than or equal to the number of processors in each dimension, it
is important to be able to keep the processor grid of roughly equal
size in each dimension.

\subsection{QCDSP Architecture:  Processor Nodes}

In 1993, to build a Teraflops machine for a few million dollars
required an inexpensive processing node with very low power
consumption.  Digital Signal Processors (DSPs) are commercial floating
point chips which are used in devices with these constraints and in
1993 were expected to provide 1\$/Megaflops performance within 2 years.
At that time the first Pentium and DEC Alpha processors were available,
with better performance per chip, but much larger cost per Megaflops
and power consumption.  Today, 670 Megaflops DSPs are available, along
with 1 Gigaflops Alpha processors, but they still differ widely in cost
and power consumption.

DSPs allow very dense packing, due to their low power.  Since memory
bandwidth was (and still is) always a problem for microprocessor based
machines, using processors, like the DSP, with relatively modest
performance made for a better match with existing DRAM speeds.  In
addition, many high-performance processors are not single chips, but
rather chip sets, where cache and a memory controller are commonly
separate chips from the processors.  Without the full compliment of
chips high-performance microprocessors can perform very modestly.

\begin{figure}[htb]
\epsfxsize=\hsize
\epsfbox{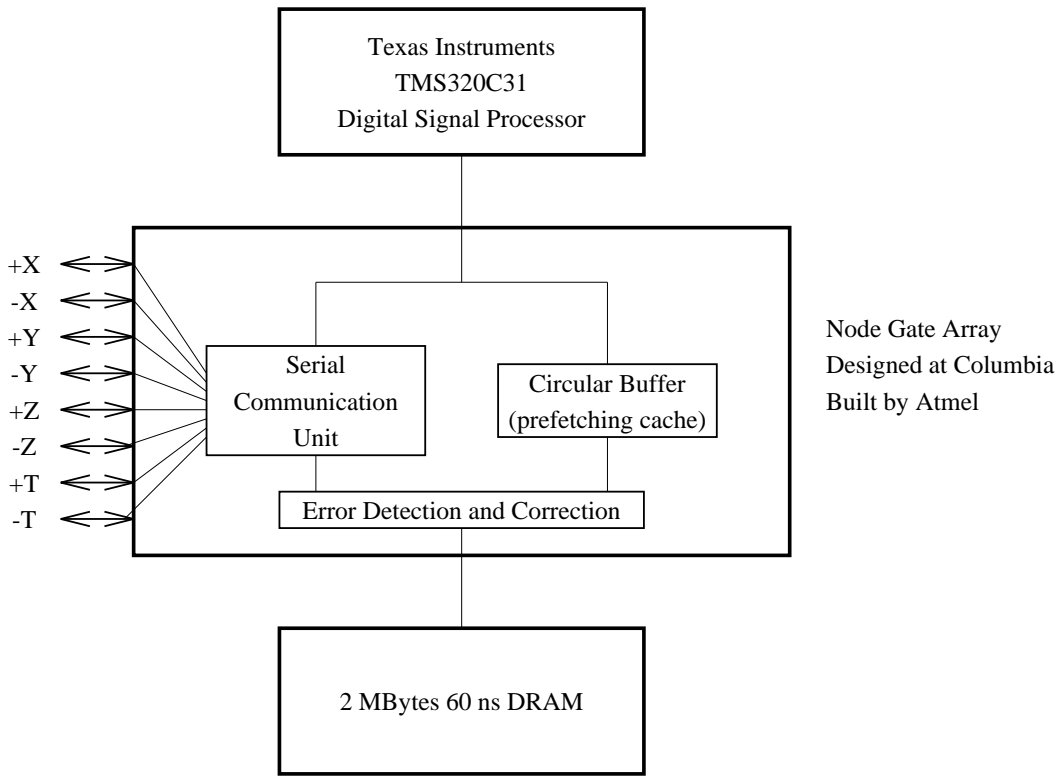}
\caption{Block diagram of a processing node of QCDSP.}
\label{fig:node_block_diag}
\end{figure}

These general considerations led to the processor nodes diagramed in
Figure \ref{fig:node_block_diag}.  This node contains a Texas
Instruments (TI) Digital Signal Processor (DSP), 2 Mbytes of EDC
corrected DRAM and a custom Application Specific Integrated Circuit
(ASIC), which we call the Node Gate Array (NGA).  This processing node
is about the size of a credit card, has a peak speed of 50 Mflops,
costs \$68 in quantity in 1997 and uses about 3 watts of total power.
By current standards, the DRAM size is small.  However, DRAM prices
were very firm during our research and development cycle and at the
prices available then, 2 MBytes of DRAM was close to 40\% of the total
parts cost.  We now describe the DSP and NGA in some detail.

\subsubsection{DSP Description}

The original DSPs began as general CPU's with only integer arithmetic
capability and found use as generally programmable controllers.  As
they achieved fixed point and then floating point capabilities, they
became useful for any application requiring fast arithmetic capability
for low cost.  They are currently used in cell phones, modems,
microwave ovens, stereo equipment, etc., anywhere that low cost, low
electrical power floating point intensive algorithms are implemented.

The 50 Mflops TI TMS320C31 DSP we use in QCDSP cost about \$50 in 1995
(\$38 in 1997) and uses about 1 watt of power.  It is a single
precision processor where double precision can be done in software with
a large performance penalty. The low power consumption makes it
possible to pack the processors close together and the cooling system
for the computer need not be more than air circulated through the
machine and through radiators fed with chilled water.  Also the entire
power bill for a 10,000 processor machine is in the range of \$50,000
per year.

DSPs generally have a smaller number of internal registers than a
conventional microprocessor and at the time QCDSP was being built, they
only contained single arithmetic units without the complicated
conditional scheduling common in high end microprocessors.  On the C31
we use, the small number of registers is offset by the presence of 2
kilowords of on chip memory.  While not identical to additional
registers, this memory can be accessed by the CPU without any delay and
is vital to getting high performance for QCD from this DSP.

From a programming point of view, writing programs (say in C or C++)
for the DSP is identical to writing for a microprocessor.  The
limitations in the DSP (small register set, modest speed) appear as
parts of the code which perform relatively more slowly, not as something
which cannot be done.  For straightforward floating point applications,
which is the dominant time used in QCD applications, the DSP performs
very well.  To support a multi-user operating system, like UNIX, the
DSP would have to do substantial swapping to memory every time a
different user input was received, due to the small number of internal
registers.

\subsubsection{NGA Description}

The NGA is the only custom integrated circuit in QCDSP, all other
components are standard commercial products (although we do have a few
Programmable Array Logic (PAL) chips which are programmed for specific
QCDSP tasks).  The NGA has a look-ahead cache (called the circular
buffer), EDC circuitry and controllers to handle the physical transfers
to the eight nearest neighbor processors in our four dimensional grid.
The NGA is described in \cite{rdm_lat94} \cite{pavlos_lat95}.

Achieving high sustained bandwidth to memory is a major difficulty in
all microprocessor based computers.  QCD calculations make this problem
somewhat easier, since for the most floating point intensive part of
the calculation, the pattern of memory fetching is regular and each
fetched floating point number is used twice.  Also, the number of words
written to memory is much smaller than the number of words read
(generally 25\% or less).  This results from the dominant calculation
involving multiplication of a complex
$3 \times 3$ matrix times a complex 3-vector.
\begin{equation}
  R_{i} = \sum_{j = 0}^2 M_{ij} \cdot V_j
\end{equation}

One can easily see that every real number on the right is used twice in
calculating the result on the left, since this is complex arithmetic.
Thus even though the DSP can only fetch one real number every 25 MHz
cycle, and it needs two per cycle to run at full speed, we can still
achieve a large fraction of peak speed through a strategy we now
describe.
\begin{enumerate}
\item  Locate the program which does $M \times V$ in on-chip memory.
\item  Copy the vector $V$ into on-chip memory.
\item  Do the multiply, fetching one code word from on-chip memory,
  one operand from on-chip memory and one operand from DRAM every
  machine cycle.
\item  Write the result out to DRAM.
\end{enumerate}
This strategy yields over 40\% of peak speed, assuming the program is
pre-loaded into on-chip memory.  Note that it makes extensive use of
the on-chip memory and assumes that DRAM can provide one operand per
cycle.  This one operand per cycle capability is made possible by the
circular buffer.

The circular buffer is a 32 word deep cache, which is given rules about
the next transfer by loading a register.  For the example above, the
circular buffer is set to fetch a maximum of 18 words from DRAM,
starting at the first address that is read (the address of $M_{00}$ in
this case).  The circular buffer is also told that subsequent addresses
accessed by the DSP will never skip by more than 2 words for this
transfer.  Thus once the first fetch of $M_{00}$ is made, the circular
buffer will immediately begin getting the remaining 17 words from
DRAM.  The circular buffer will provide these to the DSP without delay,
so the full input bandwidth can be achieved.  Since the circular buffer
stores all 18 words internally, the user can also jump back to
previously fetched words, a vital feature when the multiply requires
the Hermitian conjugate of a matrix.

The NGA also implements the four-dimensional nearest neighbor
communications network in a subsystem called the SCU (Serial Control
Unit).  The only part of the network that is not in the NGA are the
wires connecting neighboring nodes and the transceivers which are used
to drive wires which leave a motherboard.  The four-dimensional network
does physical transfers over a bit-serial connection that runs at 25 or
50 MHz.  The 50 MHz connections have proved to be stable and reliable
and are used primarily.  The SCU does automatic hardware resends when
single bit parity errors are detected.

The SCU has direct access to DRAM, without going through the circular
buffer and requires only 2 registers to be loaded to start a transfer.
This allows for very low startup latency in communications.  Users only
specify the starting address in DRAM for a transfer and the total
number of words (which can be divided into blocks with a fixed
stride).  The conversion of 32 bit words into a bit-serial stream is
handled by the SCU.

Each link between two processors runs independently of all other links
in the machine; no global synchronicity is needed or achieved.  The two
processors at each end of the link must understand which one sends and
which one receives, which for lattice QCD is trivially implemented by a
shift left/shift right approach to most data transfers.  However, the
hardware supports more generally, asynchronous message passing over the
nearest neighbor grid and one general message passing scheme has been
implemented \cite{creutz}.

QCD also requires efficient calculation of global sums across the
entire machine (for example, to know the dot product of two vectors
distributed over the entire machine).  The bit-serial communication
links cause a large overhead if global sums are done by sending a word
to a node and adding the received value to the local value and
iterating.  (The overhead comes since each 32-bit word must be entirely
received by a neighboring node before that node can use its DSP to
perform the sum.)  To avoid this, the SCU can do a global sum by adding
together, as each bit arrives, the data coming in on any set of the
communications links and from local memory.  This bit-wise sum is then
sent out over a selected wire to another node which repeats the
process.  By choosing an appropriate tree path through the machine, a
single node holds the required global sum, which it broadcasts (also
handle by the SCU with small latency) to all other nodes.

\subsection{QCDSP Architecture:  Motherboards}

For a computer with such a large number of nodes, simplicity and ease
of repair are very important.  To meet these requirements, the node
diagramed in Figure \ref{fig:node_block_diag} is contained on a single
printed circuit board (called a daughterboard). The daughterboards are
attached to the motherboards with standard 40-pin SIMM connectors, just
as DRAM is generally connected to a PC \cite{nhc} \cite{rdm_lat96}.

Each motherboard holds 63 daughterboards and a 64th processor node is
soldered to the motherboard.  This 64th processor (called node 0 on a
motherboard) is attached to an NGA, 8 MBytes of DRAM, two SCSI buses,
an EPROM, a DSP serial connection to each of the 63 daughterboard DSPs
and other electronics which controls low-level setup parameters for the
motherboard.  The 64 processors are arranged in a $4 \times 4 \times 2
\times 2 $ processor mesh.

\begin{figure}[htb]
\epsfxsize=\hsize
\epsfbox{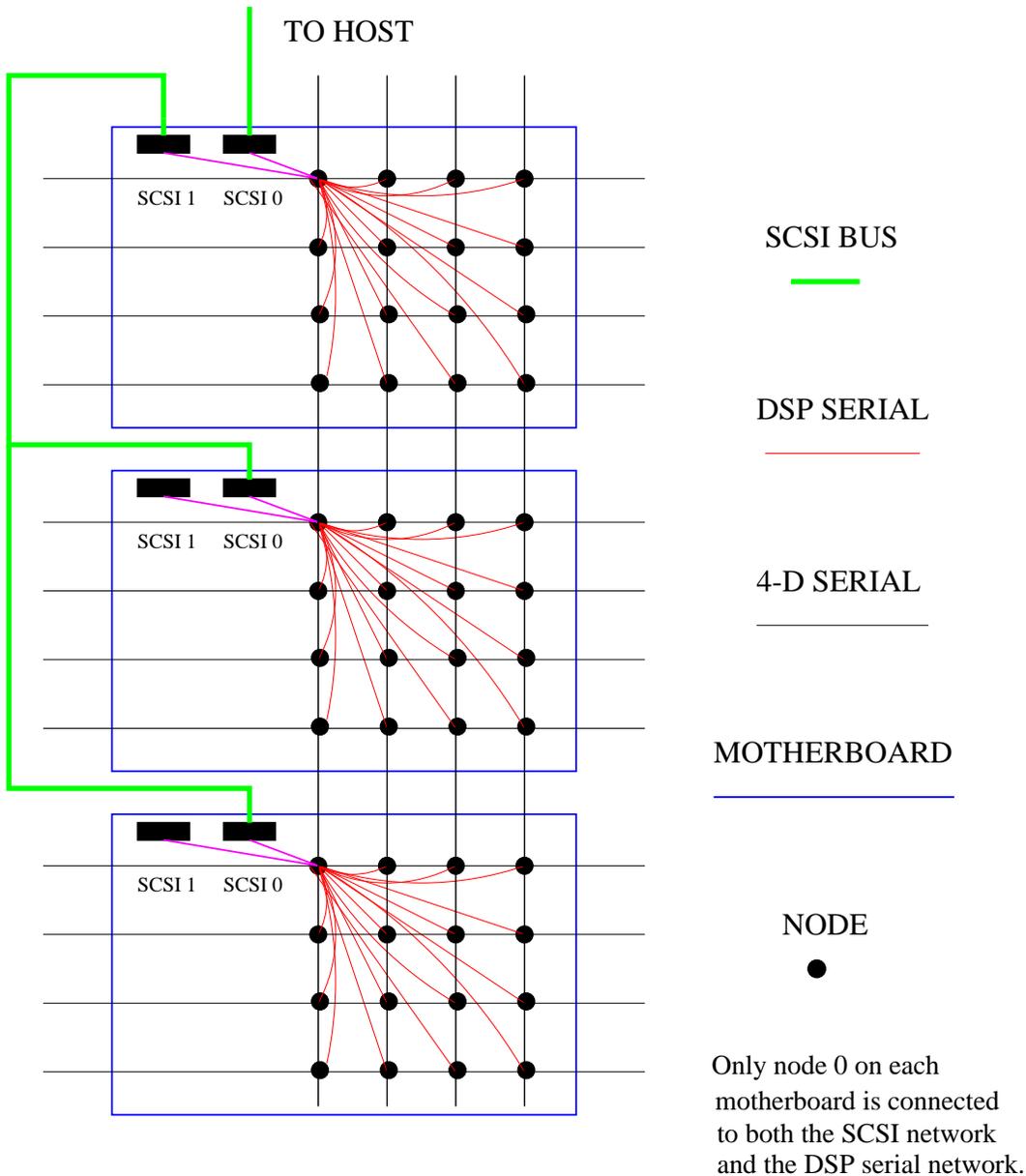}
\caption{Networks for a two-dimensional slice of QCDSP, where 16
  processing nodes are shown for a motherboard.}
\label{fig:machine_networks}
\end{figure}

Node 0 on each motherboard plays a special role during booting of
QCDSP and when I/O is done to the host workstation.  The motherboards
are connected to each other through a tree made out of the SCSI
bus connections.  Figure \ref{fig:machine_networks} shows the
configuration of the networks for a two-dimensional cross-section of
the machine.
Node 0 on each motherboard is equivalent to all the others when a
production physics calculation is underway and I/O is not being done.
In addition, each node 0 can drive a standard SCSI disk, giving a large
bandwidth to disk since each SCSI bus is independent and the bandwidth
adds.

\subsection{QCDSP Architecture:  Crates and Racks}

The low power of the DSP allows for close packing of the
daughterboards, without having to use more than forced air cooling.
Eight motherboards fit into a crate, where the backplane of the crate
provides power, the clock, the reset signal, and interrupts to the
motherboards.  Three slots in the crate can be set to run motherboards
as individual machines.  Two crates (a total of 16 motherboards)
fit in a rack, which is about four feet high, two feet wide and
three feet deep.  For pictures of the hardware, please see
the Web site http://www.phys.columbia.edu/\~{ }cqft/ and links
therein.

The extent of the four dimensional processor mesh is determined
by ribbbon cables connected to each motherboard through the
backplane.  One periodic dimension of extent 4 processors is completely
contained on the motherboard.  The remaining three dimensions
are connected with the external cables into a periodic processor
mesh that is an integral multiple of $4 \times 2 \times 2$ in
each dimension.  Changing the size of the machine requires recabling
and generally can be done in a few hours.

\section{QCDSP Software} 

Another important design objective was to make QCDSP programming as
straightforward as possible \cite{lat97}.  Although QCD algorithms are
quite well established, there are continual improvements (the domain
wall fermions described below are an example) and new techniques which
must be added.  Even leaving aside new techniques, implementing the
existing body of algorithms necessary for a complete QCD simulation
environment is sufficient effort that a reasonable programming
environment is necessary.

One major advantage to using a commercial processor in a custom
computer like QCDSP is having many software tools available.  TI
provides both an assembler and C compiler for the DSP.  In addition, we
also purchased a C++ compiler from Tartan, which has since been bought
out by TI.  The majority of lines of our programs are written in C++,
with the kernels for the floating point intensive parts written in
assembly. These are single node compilers, which do not do
parallelization for the user.  Also for the single node case, we have
commercial debuggers, evaluation modules (commercial DSP
boards hosted by a workstation or PC) and hardware emulators that we
used in code development and hardware debugging.  It is an enormous
simplification to not be responsible for developing all these tools.

QCDSP is a fully MIMD (multiple instruction, multiple data) computer.
Each processor can have a different program running on it, although
this situation requires a general communication protocol running on
each processor if inter-processor communication is required by the
programs.  For lattice gauge theory simulations, the same program is
loaded to each processor with conditional branching depending on the
processor coordinates in the four-dimensional processor grid.

\subsection{Operating System} 

The QCDSP operating system was written completely at Columbia.  Since
the memory per processor is limited, it was not possible to consider
porting Linux, for example, to QCDSP.  In addition, our application
does not require a full multitasking operating system, since whenever
the machine is doing a physics calculation, that is all it is doing.
Also, in its high performance mode, the circular buffer state is
altered by interrupts.  Therefore, a multitasking operating system
which insisted on occasionally tending to its own housekeeping chores,
would have to be switched to a non-multitasking mode during
high-performance parts of the QCD programs.  It is also much easier to
debug hardware if the software is not throwing interrupts at random
times.

In order to preserve ease of programming, the operating system provides
a UNIX-like environment to a C or C++ programmer.  Many standard C
library calls are implemented (printf, fopen, fclose, ...).  (We have
not implemented the C++ iostreams system at this time.) In addition,
there are system calls to functions which return the grid coordinates
for a processor, check the hardware status of the local processor,
return the machine size, etc.  Other system calls handle data transfers
over the nearest neighbor network.  We did not implement MPI (message
passing interface), since our hardware directly supports only a subset
of MPI calls.  It would not be very difficult to port MPI to QCDSP.

There are two major components to the operating system.  One part
resides on the host SUN workstation and other runs on QCDSP.  When
QCDSP is booted, the host begins to query the machine and determines
the total number of motherboards, daughterboards and the
four-dimensional configuration of the machine.  Recabling the machine
does not require any software changes; the new configuration is
determined on the subsequent boot.  During this process, the host
builds the appropriate tables which it uses to route information to any
particular node.

The operating system also contains a number of features for testing and
locating faulty hardware.  While booting QCDSP the host does various
hardware tests to determine whether there are any problems with the
machine.  In particular, the QCDSP run-time kernels are not loaded
until the local nodes have passed a DRAM test.  It is very important
that the operating system be able to do a substantial amount of
diagnostic work automatically on a machine with so many nodes.

When user code begins executing, the host workstation becomes a slave
to QCDSP, providing services as QCDSP requests them.  Currently, users
have access to the host file system from QCDSP and can send output to
the host console from QCDSP.  More features are planned, including the
QCDSP disk system.  During program execution, system calls can be made
to determine whether any hardware errors have occurred (parity errors on
SCU transfers, single or double bit errors in data read from DRAM).  At
the end of user program execution, the operating system scans the
machine to check the hardware state.

The operating system currently uses about 1/4 of the
available memory per node.  About half of this is used for
buffers to store the operating system log and the results of
users printf(...) calls on each node.  Users can retrieve
detailed information about their program from each node by
retrieving the print buffer contents after program completion.

\subsection{Application Software} 

Over the last several years, a large lattice QCD software package for
QCDSP has been written in C++ and assembly.  The vast majority of code
is in C++, with the implementation of the various lattice Dirac
operators written in assembly, along with SU(3) matrix and vector
routines.  Interprocessor communication is done by a set of library
routines which handle the normal transfers required in lattice QCD.

While QCDSP was being built, programs to solve the Dirac equation for
Wilson and staggered fermions were written.  These programs were used
to test the NGA design and are used as tests on the silicon wafers
when NGAs are made.  Our Wilson and staggered fermion inverters sustain
between 20 and 30\% of peak speed depending on the local volume.
Now that we have a quite complete implementation of algorithms for
lattice QCD, we plan to spend more time on the Dirac equation
solvers.  Performance between 40 and 50\% is achievable.

C++ has proved very useful for organization of this fairly large
software system.  The C++ class structure is used, for example, to
guarantee that the correct conjugate gradient solver is called for the
kind of fermions you are currently working with.  With around
10 collaborators working on the Columbia QCDSP machine and these
10 plus another 8 working on the RIKEN-BNL QCDSP machine, we
need software organization to be able to effectively share code
written by others.

Generic C++ code runs at the few percent level on QCDSP.  This is
primarily due to low memory bandwidth when program instructions and
data are being accessed from different areas of memory.  These kinds of
accesses are slowed by the delays suffered when one changes DRAM
pages.  Function calls are also slow for similar reasons, since
pushing(popping) register contents onto the stack requires
writing(reading) to(from) DRAM.

QCDSP is a general computer that has been optimized for lattice QCD.
There should be other grid based problems which would work well on this
architecture.  A completely different physics calculation has been done
on QCDSP \cite{creutz_fermions} and other applications are being
considered.

\section{QCDSP Status} 

The QCDSP machine at Columbia was finished in April, 1998.  It has now
been running production physics calculations for about a year.  During
the first few months of running, we removed processors which logged
occasional hardware errors (primarily parity errors on SCU transfers or
DRAM access).  There were also nodes which would cause the machine to
hang.  Since the communication between each node is independent of all
others, if one communication transaction does not complete, eventually
all other processors will stop as the effects of the one frozen link
cause successive neighbors to stall waiting for data to arrive.  These
kinds of errors are generally tackled by keeping a running log of the
state of each link in DRAM, so that after a hang the offending link can
be determined.

The RIKEN-BNL 0.6 Teraflops QCDSP machine was completed in October, 1998
and most of it has been in production running since then.  We are
finding the burn in time for this machine comparable to the QCDSP
machine at Columbia.  We expect the entire machine to be running
production physics very soon.  This machine was awared the Gordon
Bell prize in the performance per dollar category at SC '98 in
Orlando, Florida.

\section{Domain Wall Fermion Physics from QCDSP} 

As mentioned above, during the development of QCDSP a new lattice Dirac
operator was developed, the domain wall fermion operator.  The original
idea was due to Kaplan \cite{kaplan} and was pursued by
Shamir \cite{shamir} and Neuberger and Narayanan \cite{n-n}.  Here
we discuss the boundary variant of domain wall fermions due to
Shamir.

When the continuum Dirac operator is discretized, one can easily change
some of its properties.  In particular, until recently, all known
discretizations destroyed the chiral symmetry of the Dirac operator.
(The chiral symmetries return as the lattice spacing is taken small,
provided the parameters are adjusted appropriately.) In the continuum,
chiral symmetry says that the Dirac operator does not couple left- and
right-handed quarks to each other.  (Handedness refers to whether the
spin and momentum are parallel or antiparallel.)  For massless quarks
in the continuum, the left- and right- handed components only couple
through the dynamics of QCD, a process known as chiral symmetry
breaking.

If the discretized Dirac operator breaks chiral symmetry, then it is
hard to separate the chiral symmetry breaking due to QCD and that due
to the discretization.  Chiral symmetry breaking is one of the dominant
characteristics of the theory at low energies and it is important to
have its effects clearly represented.  Domain wall fermions are a
discretization which preserves the chiral symmetry of the theory at
finite lattice spacing,

Domain wall fermions employ a five-dimensional fermionic field, coupled
to the four-dimensional gauge (gluon) field.  The boundary conditions at
each end of the fifth dimension are chosen so that a surface state (a
mode that propagates in four dimensions) appears which is chiral.  In
particular, a right-handed, four dimensional quark appears at one end of
the fifth dimension and a left-handed four dimensional quark at the
other.  These states are the four dimensional chiral quarks we desire.
As the extent of the fifth dimension, $L_s$, is taken large, the domain
wall Dirac operator breaks chiral symmetry with terms of order
$\exp(-\alpha L_s)$ where $\alpha$ is a constant.

Computationally, domain wall fermions cost a factor of $L_s$ more than
other approaches.  Simulations done by the Columbia group and others
show that for smaller lattice spacing, $L_s = 16$ is likely sufficient,
while at larger lattice spacing, $L_s = 32 $ or more may be necessary.
How much one gains from having chiral symmetry must be balanced against
this additional cost.

\subsection{QCD Thermodynamics}

When QCD is heated up, the quarks and gluons which compose hadronic
matter are liberated into a quark-gluon plasma.  Lattice simulations
have found this temperature to be $\sim 160$ MeV.  However, the
detailed properties of the phase transition are expected to be
controlled by the symmetries of the theory, including the chiral
symmetries.  Since domain wall fermions have the correct chiral
symmetries even at finite lattice spacing, it is important to see if
our understanding of the critical region of the QCD phase transition
changes with this formulation.

\begin{figure}[htb]
\epsfxsize=\hsize
\epsfbox{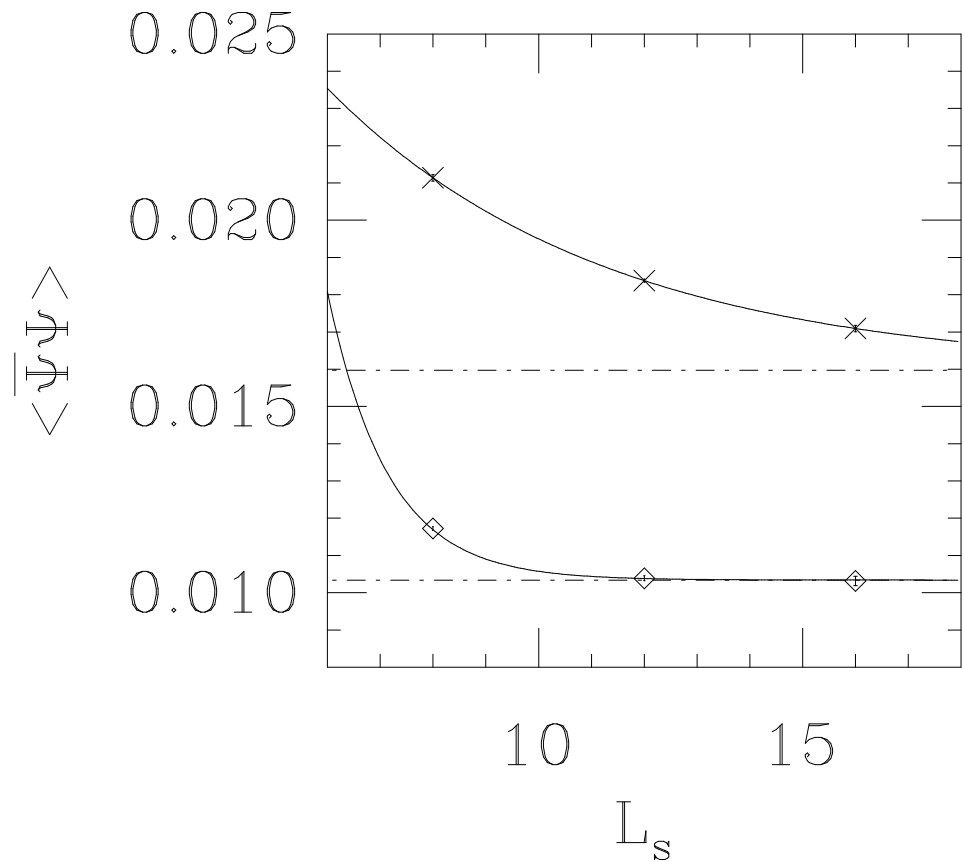}
\caption{Dependence of the chiral condensate versus $L_s$ for
  $8^3 \times 4$ lattices above (lower curve) and below (upper curve)
  the critical temperature.}
\label{fig:pbp_vs_ls}
\end{figure}

The group at Columbia has been actively studying the finite temperature
QCD phase transition with domain wall fermions using
QCDSP \cite{thermo}.  At the large lattice spacings where current
thermodynamics studies can be done, large values for $L_s$ are
required.  Figure \ref{fig:pbp_vs_ls} shows the dependence on the
length of the fifth dimension of the chiral condensate.  (The chiral
condensate should go to zero with the quark mass in the quark-gluon
phase and be non-zero in the normal hadronic phase.)  One can
see the expected exponential fall-off for the $L_s$ dependent
effects.
%
%

We have done first simulations of the phase transition region for QCD
using domain wall fermions and $L_s = 24$.  This is not a large enough
value for $L_s$ to completely remove the exponentially small chiral
symmetry breaking effects.  However, we did find the temperature for
the QCD phase transition with domain wall fermions to be $\sim 170$
MeV, which is consistent with other techniques.  We are currently doing
more simulations with smaller chiral symmetry breaking effects.  The
power of the QCDSP computers is vital for these studies.

\subsection{Weak Matrix Elements}

Part of testing the standard model of particle physics involves knowing
precisely the effects of weak interactions on hadronic states.  We must
use computational techniques to make the hadrons and then measure the
weak interaction effects in these hadrons made on the computer.  The
process of inserting the weak interaction effects into the hadrons is
much more controlled if chiral symmetry is intact.  Without chiral
symmetry, different effects can become intermixed.  In addition, chiral
symmetry tells us that the behavior of weak interactions in certain
hadronic states becomes small as the quark mass becomes small.  Without
chiral symmetry to enforce this condition, one ends up calculating a
small number by subtracting two large numbers.  This is very costly and
introduces large errors.

In the realm of weak interactions in QCD systems, the topic of CP
violation is of particular importance.  The standard model allows the
combined symmetry of charge conjugation and parity (CP symmetry) to be
violated, due to the presence of one complex parameter in the theory.
CP violation was first measured experimentally in 1964 by Cronin and
Fitch in the kaon system (the kaon is a bound state of two quarks, one
of which is the strange quark).  This measurement of CP violation
through what is called mixing has recently been joined by a new
experimental announcement of CP violation in decays.  Without detailed
QCD calculations, one cannot know if both effects are consistent with a
single value for the complex parameter in the standard model.

The first work on weak matrix elements with domain wall fermions was
done by \cite{blum_soni} who calculated a parameter related to CP
violation by mixing for quenched QCD.  (This had previously been
done by other groups using Wilson and staggered fermions.)  They found
that for moderately small lattice spacings, an $L_s$ of $\sim 16$
effectively restored chiral symmetry for the domain wall fermions.  In
a joint work using the QCDSP computer at the RIKEN-BNL Research Center,
a collaboration of the RIKEN-BNL, BNL and Columbia lattice groups
(which includes the authors of \cite{blum_soni}) are using domain wall
fermions to calculate the CP violation in mixing and decays, for
quenched QCD.

CP violation in decays has been worked on for some time using other
fermions formulations.  The lack of the full symmetries of the theory
has made these calculations very difficult, a problem which is solved
by the domain wall fermions.  The domain wall fermions make the
calculation many times more computationally expensive, but may solve
enough other problemss that a final answer is possible.  We are
anxiously awaiting the completion of this calculation.

\section{Conclusions}

The QCDSP computer is a very cost effective computer for calculations
in lattice QCD.  This computer was designed and constructed by a small
group of people, primarily physicists, over five years with a total
parts cost of $\sim 4$ million dollars, including research and
development.  (Salaries add an additional $\sim 1.5$ million dollars to
the cost.) The final machine has a cost performance of about
\$10/Megaflops and won the 1998 Gordon Bell prize in the cost
performance category at SC '98 in Orlando, Florida.

Including the Columbia and RIKEN-BNL computers, 20,480 processors with
a peak speed of over 1 Teraflops are available for QCD calculations.
Physicists at Columbia, BNL and RIKEN-BNL are aggressively using these
machines to study QCD, focusing on using the new domain wall fermion
formalism.  Calculations for QCD thermodynamics and weak matrix
elements, among others, are well underway.

{\bf Acknowledgements}

The QCDSP computer was developed with
funds provided by the United States Department of Energy.
Funds for the 0.4 Teraflops QCDSP computer were provided by
the US DOE, while the 0.6 Teraflops computer was funded by
the RIKEN-BNL Research Center.  The work discussed here is
the cumulative effort of many individuals over a number of years.
They are:
\begin{description}
\item[Columbia University:] \mbox{}
  \begin{description}
    \item[Current Members:]  Ping Chen, Norman Christ,
      George Fleming, Tim Klassen, Robert Mawhinney, Gabi Siegert,
      ChengZhong Sui, Pavlos Vranas, Lingling Wu and Yuri Zhestkov.
    \item[Former Members:] Igor Arsenin, Dong Chen,
      Chulwoo Jung, Adrian \newline Kaehler, Yubing Luo and Catalin
      Malureanu.
  \end{description}
\item[Columbia University Nevis Laboratories:]
  Alan Gara and John Parsons
\item[Brookhaven National Laboratory:]
  Michael Creutz, Chris Dawson and \newline Amarjit Soni.
\item[RIKEN-BNL Research Center:]
  Tom Blum, Shigemi Ohta, Shoichi Sasaki and Matthew Wingate
\item[SCRI at Florida State:] Robert Edwards and Tony Kennedy (now
  at Edinburgh)
\item[The Ohio State University:] Greg Kilcup
\item[Trinity College, Dublin:] Jim Sexton
\item[Fermilab:]  Sten Hanson
\end{description}

\end{document}